\documentclass[fontsize=9pt]{article}
\usepackage{spconf,amsmath,graphicx}
\usepackage{amssymb}
\usepackage{booktabs}
\usepackage{amsthm}
\usepackage{tabularx}
\usepackage{caption}
\usepackage{subcaption}
\usepackage{multirow}
\usepackage{diagbox}
\usepackage[breaklinks=true,bookmarks=false]{hyperref}
\usepackage{enumitem}
\usepackage{mathtools}
\usepackage{makecell}
\usepackage{bbm,color}
\usepackage{algorithm,algorithmic}
\usepackage{multirow}
\newcommand{\etal}{\textit{et al.}}
\newtheorem{mythm}{Theorem}

\newtheorem{lemma}[mythm]{Lemma}

\DeclareMathOperator*{\argmax}{arg\,max}
\DeclareMathOperator{\sign}{sign}

\DeclareMathOperator*{\argmin}{arg\,min}

\title{Weighted Anisotropic -- Isotropic Total Variation for Poisson Denoising
}

\name{$\textnormal{Kevin Bui}^1$, $\textnormal{Yifei Lou}^2$, $\textnormal{Fredrick Park}^3$, and $\textnormal{Jack Xin}^1$
\thanks{The work was partially supported by NSF grants DMS-1846690, DMS-1854434, DMS-1952644, DMS-2151235, and a Qualcomm Faculty Award.}}
\address{
$^{1}$Department of Mathematics;
University of California, Irvine;
Irvine, CA 92697, United States \\
$^{2}$Department of Mathematical Sciences;
University of Texas, Dallas; Richardson, TX 75080, United States\\
$^{3}$Department of Mathematics \& Computer Science;
Whittier College; Whittier, CA 90602, United States
}

\begin{document}
%
\maketitle
\begin{abstract}
Poisson noise commonly occurs in images captured by photon-limited imaging systems  such as in astronomy and medicine. As the distribution of Poisson noise depends on the pixel intensity value,  noise levels vary from pixels to pixels. Hence, denoising a Poisson-corrupted image while preserving important details can be challenging. In this paper, we propose a Poisson denoising model by incorporating the weighted anisotropic--isotropic total variation (AITV) as a regularization. We then develop an alternating direction method of multipliers with a combination of a proximal operator for an efficient implementation. Lastly, numerical experiments demonstrate that our algorithm outperforms other Poisson denoising methods in terms of image quality and computational efficiency.
\end{abstract}


\section{Introduction}
In various applications such as astronomy \cite{lanteri2005restoration} and medicine \cite{vardi1985statistical}, photon-counting devices are utilized to capture images. However, these images are susceptible to Poisson noise, in which any measured  intensity at each pixel is a realization of a Poisson random variable with mean and variance equal to its true value. Recall that a Poisson random variable with mean and variance has a probability distribution function
$\mathbb{P}_{\mu}(n) = \frac{e^{-\mu}\mu^n}{n!},\; n \geq 0. $
Let $g$ be a clean image of size $M \times N$. If $g$ is corrupted by Poisson noise, then its corresponding noisy measurement $f$ can be formulated as
\begin{align*}
    f_{i,j} \sim \text{Poisson}(g_{i,j}),\; 1 \leq i \leq M, 1 \leq j \leq N.  
\end{align*}

One general approach for Poisson denoising is by the maximum a posteriori (MAP)
\begin{align*}
    \prod_{i,j} \mathbb{P}(u_{i,j}|f_{i,j})  = \prod_{i,j} \frac{e^{-u_{i,j}} u_{i,j}^{f_{i,j}}}{(f_{i,j})!} \frac{\mathbb{P}(u_{i,j})}{\mathbb{P}(f_{i,j})},
\end{align*}
 with respect to an estimated image $u$ from a noisy input $f.$
Taking the negative logarithm yields the following optimization problem to minimize:
\begin{align*}
    \sum_{i,j} u_{i,j} - f_{i,j} \log u_{i,j} - \log \mathbb{P}(u_{i,j}).
\end{align*}
The  term $- \log \mathbb{P}(u_{i,j})$ can be regarded as an image prior. A classic choice is the total variation (TV)~\cite{rudin1992nonlinear}. Le \etal \  \cite{le2007variational} derived a TV-regularized model  for Poisson denoising:
\begin{align} \label{eq:isotropic_poisson_model}
    \min_{u} \lambda \langle u - f \log u, \mathbbm{1} \rangle + \|\nabla u \|_{2,1}, 
\end{align}
where $\mathbbm{1}$ denotes the all-one vector and $\|\nabla u \|_{2,1}$ is the isotropic form of TV. TV has been a popular regularizer for denoising, but it only performs well on piecewise-smooth images. For natural images, TV tends to introduce staircasing artifacts, lose image contrasts, and smear out fine details \cite{lysaker2003noise}. Several  variants of TV have been proposed including nonlocal TV \cite{holla2018non} and total generalized variation \cite{wang2013iterative}. Unfortunately, these TV variants are numerically complicated to compute. As an alternative, fractional-order total variation (FOTV) has a convenient, compact discrete form and was demonstrated to perform well in Poisson denoising \cite{rahman2020poisson}. 

More complex models such as nonlocal methods \cite{kumar2019low, salmon2014poisson, zha2019rank, zhao2021nonlocal, zha2022simultaneous, zha2020image} and convolutional neural networks (CNNs) \cite{feng2017fast, remez2018class, ren2018dn} have been developed to perform Poisson denoising.  However, nonlocal methods are computationally expensive due to the needs to compare image patches in groups in terms of their similarity and to enforce a low-rank structure.  As highly nonconvex models, CNNs require an adequately large training set, hyperparameter tuning (e.g., batch size, learning rate, number of layers, etc.), and heavy computational resources such as GPUs. Since CNNs demonstrate the promising potential of nonconvex modeling, one direction is to develop a nonconvex variant of \eqref{eq:isotropic_poisson_model} to avoid most of their computational limitations. 

Studies have demonstrated that nonconvex regularizers can preserve edges better than convex models
\cite{zeng2018edge}. For example, TV$^p (0 <p< 1)$ preserves edges better than the convex TV model in Gaussian denoising \cite{chen2012non}. Another nonconvex TV variant is called weighted anisotropic--isotropic TV (AITV) \cite{lou2015weighted} that recovers sharper images than TV and TV$^p$ for images corrupted by Gaussian noise \cite{lou2015weighted}. Despite its recent success, AITV has not yet been applied in Poisson denoising and is slow in computations. In this paper, we propose a variational Poisson denoising model with the AITV regularization and improve the efficiency by taking advantage of a proximal operator \cite{lou2018fast} over its original implementation that uses the difference-of-convex algorithm (DCA).


\section{The Proposed Approach} \label{sec:proposed_model}
\subsection{Model Formulation}
Two  popular numerical approximations of TV are the isotropic TV and the anisotropic TV. Specifically, we represent an image as an $M \times N$ matrix and define an Euclidean space  $X \coloneqq \mathbb{R}^{M \times N}$ with the standard  inner product $\langle \cdot, \cdot \rangle_X$ and the Euclidean norm $\|\cdot\|_2$. We will omit the subscript $X$ and  use $\langle \cdot, \cdot \rangle$ for the sake of brevity.

To discretize the image gradient, we define another Euclidean space $Y \coloneqq X \times X$. The discrete gradient operator $\nabla: X \rightarrow Y$ is given by $(\nabla u)_{i,j} = \left((\nabla_x u)_{i,j},
    (\nabla_y u)_{i,j} \right)$,
where $\nabla_x, \nabla_y$ are the horizontal and vertical difference operators.
For the space $Y$, we define the inner product by
\[
  \langle p, q \rangle = \langle p_1, q_1 \rangle_X + \langle p_2, q_2 \rangle_X,
\]
for $p=(p_1, p_2), q=(q_1, q_2) \in Y$.  We also define the following norms on $Y$:
\begin{align*}
    \|p\|_1 &= \sum_{i=1}^M \sum_{j=1}^N |(p_1)_{i,j}|+|(p_2)_{i,j}|, \\
    \|p\|_2 & = \sqrt{\sum_{i=1}^M \sum_{j=1}^N|(p_1)_{i,j}|^2+|(p_2)_{i,j}|^2} , \\
    \|p\|_{2,1} &= \sum_{i=1}^M \sum_{j=1}^N \sqrt{(p_1)_{i,j}^2 + (p_2)_{i,j}^2}. 
\end{align*}

By our definitions, the isotropic TV and anisotropic TV can be formulated as $\|\nabla u\|_{2,1}$ and $\|\nabla u\|_1,$ respectively. Unfortunately both of these fail in recovering oblique edges \cite{condat2017discrete}. 
To mitigate this artifact, Lou \etal \  \cite{lou2015weighted} proposed the AITV regularizer $\|\nabla u\|_1 - \alpha \|\nabla u\|_{2,1}$, where the parameter $\alpha\in[0,1]$ controls the sparsity of the gradient at each pixel. Replacing the isotropic TV in \eqref{eq:isotropic_poisson_model} with AITV for a pre-defined $\alpha \in [0,1]$, we arrive at the proposed model:
\begin{align}\label{eq:aitv_poisson}
    \min_{u}  \lambda \langle u - f \log u, \mathbbm{1} \rangle + \|\nabla u\|_1 - \alpha \|\nabla u\|_{2,1}.
    \end{align}

\subsection{Numerical Algorithm}
We develop an alternating direction method of multipliers (ADMM) \cite{boyd2011distributed} to solve for \eqref{eq:aitv_poisson}. By introducing two auxiliary variables $v \in X$ and $w = (w_x, w_y) \in Y,$  we have the following constrained optimization problem:
    \begin{equation}
\begin{aligned}\label{eq:constrained_opt}
    \min_{u,v, w}  & \quad  \lambda \langle v - f \log v, \mathbbm{1} \rangle +  \|w \|_1 - \alpha  \|w\|_{2,1} \\
    \text{s.t.} & \quad u = v \quad\mbox{and} \quad \nabla u = w.
\end{aligned}
\end{equation}
Then its augmented Lagrangian is written as
\begin{align*}
    &\mathcal{L}_{\beta}(u, v,w, y, z) =  \lambda \langle v - f \log v, \mathbbm{1} \rangle + \|w \|_1 - \alpha  \|w\|_{2,1}\\ &+ \langle y, u -v \rangle + \frac{\beta}{2} \|u - v\|_2^2+ \langle z, \nabla u -w \rangle + \frac{\beta}{2} \|\nabla u - w\|_2^2,
        \end{align*}
where $y \in X, \ z = (z_x, z_y) \in Y$ are the Lagrange multipliers and $\beta > 0$ is a penalty parameter. As a result,   ADMM iterates as follows:
\begin{subequations}
    \begin{align}
    u_{k+1} &= \argmin_u \mathcal{L}_{\beta} (u, v_k, w_k, y_k, z_k) \label{eq:u_subprob}\\
    v_{k+1} &= \argmin_v \mathcal{L}_{\beta} (u_{k+1}, v, w_k, y_k, z_k) \label{eq:v_subprob}\\
    w_{k+1} &= \argmin_w \mathcal{L}_{\beta} (u_{k+1}, v_{k+1}, w, y_k, z_k) \label{eq:w_subprob}\\
    y_{k+1} &= y_k + \beta_k(u_{k+1} - v_{k+1}) \label{eq:y_eq}\\
    z_{k+1} &= z_k + \beta_k(\nabla u_{k+1} - w_{k+1}) \label{eq:z_eq}\\
    \beta_{k+1}  &= \sigma \beta_k, \label{eq:beta_update}
    \end{align}
\end{subequations}
where $\sigma > 1$. The last step \eqref{eq:beta_update} is inspired from \cite{gu2017weighted} to accelerate the numerical convergence of  ADMM. If $\sigma$ is too large, the algorithm might stop too early, yielding an unsatisfactory solution. Hence, $\sigma$ needs to be chosen carefully.

We derive closed-form solutions for the subproblems \eqref{eq:u_subprob}-\eqref{eq:w_subprob}. The first-order optimality condition for \eqref{eq:u_subprob}  is
\begin{align} \label{eq:first_order_u}
    \beta_k (I - \Delta ) u_{k+1} = \beta_{k}v_k - y_k - \nabla^{\top}(z_k - \beta_{k}w_k),
\end{align}
where $\Delta = -\nabla^{\top}\nabla$ is the Laplacian operator. By assuming periodic boundary condition for $u$, \eqref{eq:first_order_u} can be solved efficiently by the 2D discrete Fourier transform $\mathcal{F}$~\cite{wang2008new}, thus leading to an update of $u_{k+1}$ to be
\begin{align}\label{eq:u_update}
    u_{k+1} = \mathcal{F}^{-1}\left( \frac{\mathcal{F}(\beta_kv_k -y_k) - \mathcal{F}(\nabla)^* \circ \mathcal{F}(z_k - \beta_k w_k)}{\beta_k \mathcal{F}(I-\Delta)} \right),
\end{align}
where $\mathcal{F}^{-1}$ is the inverse Fourier transform, the superscript $*$ denotes complex conjugate,  $\circ$ denotes the componentwise product,  and the division is componentwise as well. By taking derivative of \eqref{eq:v_subprob} with respect to $v$ and setting it to zero, 
 we get the closed-form solution for
\begin{align}\label{eq:v_update}
    v_{k+1} = \frac{r_k+\sqrt{r_k^2+4\lambda\beta_k f}}{2\beta_k},
\end{align}
where $r_k = \beta_k u_{k+1} +y_{k} - \lambda \mathbbm{1}$ and all the operations (square root, square, and division) are  componentwise. Lastly, the $w$-subproblem \eqref{eq:w_subprob} can be decomposed independently at each pixel $(i,j),$ i.e.,
\begin{gather}
\begin{aligned}\label{eq:w_update}
    (w_{i,j})_{k+1} &=\argmin_{w_{i,j}}  \|w_{i,j}\|_1 - \alpha \|w_{i,j}\|_2\\  
    &+ \frac{\beta_k}{2}  \left\| w_{i,j} - \left( (\nabla u_{k+1})_{i,j} + \frac{(z_k)_{i,j}}{\beta_k}\right) \right\|_2^2. 
\end{aligned}
\end{gather}
The optimization problem for each component of $w$ is a special case of the proximal operator for $\ell_1 - \alpha \ell_2$, defined by
\begin{align}\label{eq:prox}
    \text{prox}(x, \alpha, \beta) =\argmin_y \|y\|_1 - \alpha \|y\|_2 + \frac{\|x-y\|_2^2}{2 \beta}. 
\end{align}
With the help of the proximal operator \eqref{eq:prox}, we obtain a closed-form solution to update every $w_{i,j}$ by
\[
  (w_{i,j})_{k+1}= \text{prox}\left((\nabla u_{k+1})_{i,j} + \frac{(z_k)_{i,j}}{\beta_k}, \alpha, \frac{1}{\beta_k}\right).
\]
As derived in \cite{lou2018fast}, the proximal operator for $\ell_1 - \alpha \ell_2$ has a closed-form solution formulated in Lemma \ref{lemma:prox}.

\begin{lemma}[\cite{lou2018fast}]\label{lemma:prox}
Given $x \in \mathbb{R}^n$, $\beta >0$, and $\alpha \in [0,1]$, the optimal solution to \eqref{eq:prox} is given by one of the following cases:
\begin{enumerate}
    \item When $\|x\|_{\infty} > \beta$, we have
        $x^* = (\|\xi\|_2 + \alpha \beta) \frac{\xi}{\|\xi\|_2}$,
     where $\xi= \sign(x)\circ\max(|x|-\beta,0)$. 
    \item When $(1-\alpha) \beta < \|x\|_{\infty} \leq \beta$, then $x^*$ is a 1-sparse vector such that one chooses $i \in \displaystyle \argmax_j(|x_j|)$ to define $x^*_i=\left(|x_i| + (\alpha-1)\beta\right)\sign(x_i)$ and set the remaining  elements  to 0.
\item When $\|x\|_{\infty} \leq (1- \alpha)\beta$, then $x^* = 0$. 
\end{enumerate}
\end{lemma}

The overall ADMM framework to solve \eqref{eq:aitv_poisson} is described in Algorithm \ref{alg:admm}. By emulating the proof of \cite[Theorem 2]{gu2017weighted}, we have $\|u_{k+1} - u_k \|_2 \rightarrow 0$, which corresponds to the stopping criterion in Algorithm \ref{alg:admm}. Algorithm \ref{alg:admm} is expected to converge within a reasonable number of iterations. Although global convergence was proven for nonconvex ADMM \cite{wang2019global}, it may not be guaranteed for our algorithm since the gradient operator does not satisfy the necessary surjectivity condition.

\begin{algorithm}[h!!!]
\caption{ADMM for \eqref{eq:aitv_poisson}}
\label{alg:admm}
\scriptsize
\begin{algorithmic}[1]
\REQUIRE Noisy image $f$, fidelity parameter $\lambda$, penalty parameter $\beta_0$, penalty multiplier $\sigma > 1$.\\
    \STATE Initialize $u_0, w_0, z_0$.\\
    \STATE Set $k=0$.\\
   \WHILE{$\frac{\|u_{k}-u_{k-1}\|_2}{\|u_{k}\|_2} > \epsilon$}
   \STATE Compute $u_{k+1}$ by \eqref{eq:u_update}.
   \STATE Compute $v_{k+1}$ by \eqref{eq:v_update}.
   \STATE Compute $w_{k+1}$ by \eqref{eq:w_update}.
   \STATE $y_{k+1} = y_k + \beta_k(u_{k+1} - v_{k+1})$.
   \STATE  $z_{k+1} = z_k + \beta_k(\nabla u_{k+1} - w_{k+1})$.
    \STATE $\beta_{k+1}  = \sigma \beta_k$.
    \STATE $k \coloneqq k+1$.
   \ENDWHILE
   \RETURN Denoised image $u^*=u_{k+1}$.\\
\end{algorithmic}
\end{algorithm}


\section{Numerical Results} \label{sec:experiment}
\begin{figure}
\centering
               \begin{subfigure}[b]{0.10\textwidth}
         \centering
         \includegraphics[width=\textwidth]{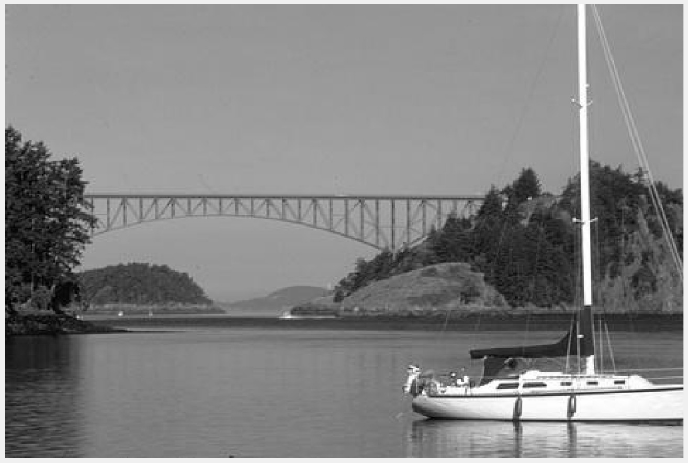}
         \caption{River}
         \label{fig:river}
     \end{subfigure}
     \begin{subfigure}[b]{0.10\textwidth}
         \centering
         \includegraphics[width=\textwidth]{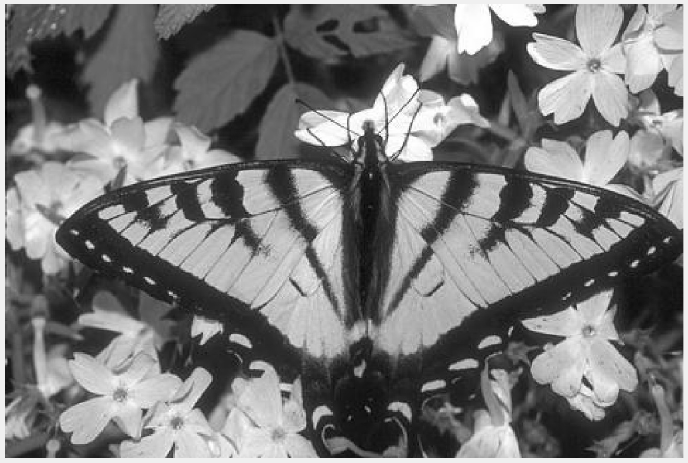}
         \caption{Butterfly}
         \label{fig:butterfly}
     \end{subfigure}
     \begin{subfigure}[b]{0.10\textwidth}
         \centering
         \includegraphics[width=\textwidth]{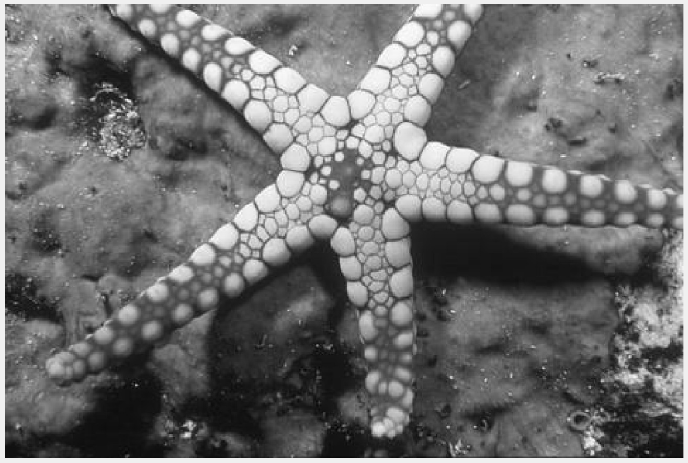}
         \caption{Starfish}
         \label{fig:starfish}
     \end{subfigure}\\
          \begin{subfigure}[b]{0.10\textwidth}
         \centering
         \includegraphics[scale=0.15]{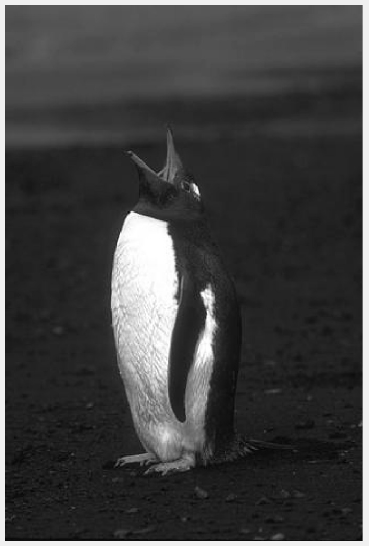}
         \caption{Penguin}
         \label{fig:penguin}
     \end{subfigure}
          \begin{subfigure}[b]{0.10\textwidth}
         \centering
         \includegraphics[scale=0.15]{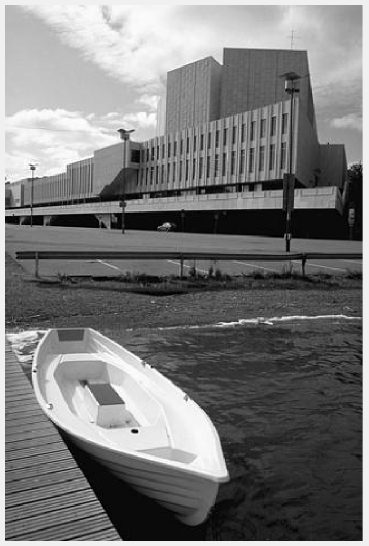}
         \caption{Boat}
         \label{fig:boat}
     \end{subfigure}
        \caption{Original natural images for testing. The image sizes are either $321 \times 481$ (top) or $481 \times 321$ (bottom).}
        \label{fig:experiment_image}
\end{figure}

We evaluate the AITV-regularized Poisson denoising model \eqref{eq:aitv_poisson} on five grayscale images selected from the Berkeley Segmentation Dataset~\cite{MartinFTM01}. The original images are shown in Figure \ref{fig:experiment_image}. We compare our proposed AITV model with the classical TV \cite{le2007variational}, non-local PCA (NL-PCA) \cite{salmon2014poisson}, and a 
recent Poisson denoising method by FOTV \cite{rahman2020poisson}. We use the MATLAB codes provided by the authors of NL-PCA and FOTV. Note that FOTV (including TV as its special case) is solved by ADMM, which is different from Algorithm \ref{alg:admm} in that its penalty parameter $\beta$ is fixed and it solves a nonlinear equation per iteration.  Quantitatively, we evaluate the performance of image denoising by peak-signal-to-noise ratio (PSNR) and structural similiarity index (SSIM). 
The experiments are performed in MATLAB R2021b on a Dell laptop with a 1.80 GHz Intel Core i7-8565U processor and 16.0 GB of RAM. The code is available at \url{https://github.com/kbui1993/Official_AITV_Poisson_Denoising}.

\begin{table}[h!]
	\caption{PSNR/SSIM of the four denoising methods under three  peak values of the original images in Figure \ref{fig:experiment_image}. \textbf{Bold} indicates the best value. The last column is the average over the five testing images.}
	\scriptsize
	\begin{center}
		\resizebox{0.485\textwidth}{!}{%
			\begin{tabular}{|c|l|c|c|c|c|c||c|} \hline
				Peak &\backslashbox{Method}{Image} & river & butterfly & starfish & penguin & boat & Avg. \\ \hline
				\multirow{ 6}{*}{80} &          Noisy & 22.03/0.36 & 22.61/0.53 & 22.73/0.56 & 24.75/0.35 & 22.46/0.53 & 22.92/0.47 \\ \cline{2-8}
				 &NL-PCA & 29.35/0.71 & 28.15/\textbf{0.83} & 28.52/\textbf{0.83} & 33.41/0.88 & 28.23/0.79 & 29.53/0.81 \\ \cline{2-8}
				&TV & 30.39/0.84 & 28.40/\textbf{0.83} & 28.31/0.81 & 33.40/0.87 & 28.40/0.80 & 29.78/0.83 \\ \cline{2-8}
        &FOTV & 30.20/0.84 & \textbf{28.50}/0.82 & \textbf{28.63}/0.82 & 33.25/0.87 & 28.23/0.78 & 29.76/0.83 \\ \cline{2-8}
        &AITV & \textbf{30.70}/\textbf{0.85} & 28.46/\textbf{0.83} & 28.33/0.81 & \textbf{33.89}/\textbf{0.90} & \textbf{28.69}/\textbf{0.81} & \textbf{30.01}/\textbf{0.84} \\ \hline \hline
        				\multirow{ 6}{*}{55} &                  Noisy & 20.38/0.30 & 20.54/0.46 & 21.08/0.48 & 23.08/0.28 & 20.81/0.47 & 21.18/0.40 \\ \cline{2-8}
        				&NL-PCA & 28.42/0.75 & \textbf{27.46}/\textbf{0.80} & \textbf{27.88}/\textbf{0.80} & 32.81/0.85 & 27.54/0.75 & 28.82/0.79 \\ \cline{2-8}
        				 &TV & 29.06/\textbf{0.83} & 27.26/\textbf{0.80} & 27.32/0.77 & 32.51/0.87 & 27.29/0.76 & 28.69/0.80 \\ \cline{2-8}
        &FOTV & 28.93/0.76 & 27.40/0.78 & 27.62/0.79 & 32.03/0.80 & 27.17/0.76 & 28.63/0.78 \\ \cline{2-8}
        &AITV & \textbf{29.67}/\textbf{0.83} & 27.43/\textbf{0.80} & 27.36/0.78 & \textbf{33.26}/\textbf{0.88} & \textbf{27.67}/\textbf{0.77} & \textbf{29.08}/\textbf{0.81} \\ \hline \hline
                				\multirow{ 6}{*}{30} &                          Noisy & 17.74/0.23 & 17.86/0.36 & 18.41/0.36 & 20.42/0.18 & 18.21/0.37 & 18.53/0.30 \\ \cline{2-8}
                				&NL-PCA & 27.57/0.68 & \textbf{25.87}/0.72 & \textbf{26.29}/\textbf{0.73} & 31.96/0.85 & 26.16/\textbf{0.72} & \textbf{27.57}/0.74 \\  \cline{2-8}
                				&TV & 27.14/0.61 & 25.46/0.70 & 25.71/0.70 & 29.68/0.71 & 25.62/0.68 & 26.72/0.68 \\ \cline{2-8}
        &FOTV & 27.11/0.68 & 25.58/0.71 & 25.94/0.72 & 28.69/0.61 & 25.51/0.68 & 26.57/0.68 \\ \cline{2-8}
        &AITV & \textbf{28.18}/\textbf{0.79} & 25.71/\textbf{0.75} & 25.79/0.72 & \textbf{31.99}/\textbf{0.87} & \textbf{26.17}/\textbf{0.72} & \textbf{27.57}/\textbf{0.77} \\ \hline
		\end{tabular}}
	\end{center}
	\label{tab:result}
\end{table}     

\begin{table}[h!]
    \centering
    \scriptsize
    \begin{tabular}{l|c}
    Method & Avg. Time (s)\\ \hline
             NL-PCA & 20.05 \\
        TV & 16.59 \\
         FOTV & 17.47 \\
         AITV & 1.89
    \end{tabular}
    \caption{Average computational time in seconds. }
    \label{tab:time}
\end{table}
\begin{figure*}
     \centering
               \begin{subfigure}[b]{0.16\textwidth}
         \centering
         \includegraphics[width=\textwidth]{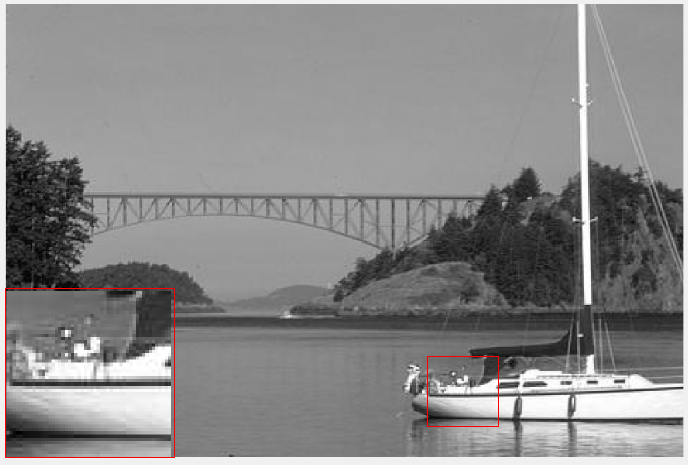}
         \caption{clean}
         \label{fig:river_clean}
     \end{subfigure}
     \begin{subfigure}[b]{0.16\textwidth}
         \centering
         \includegraphics[width=\textwidth]{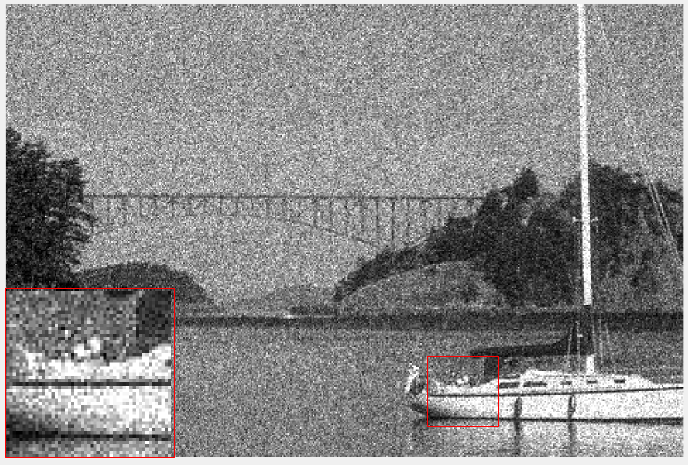}
         \caption{noisy}
         \label{fig:river_noisy}
     \end{subfigure}
     \begin{subfigure}[b]{0.16\textwidth}
         \centering
         \includegraphics[width=\textwidth]{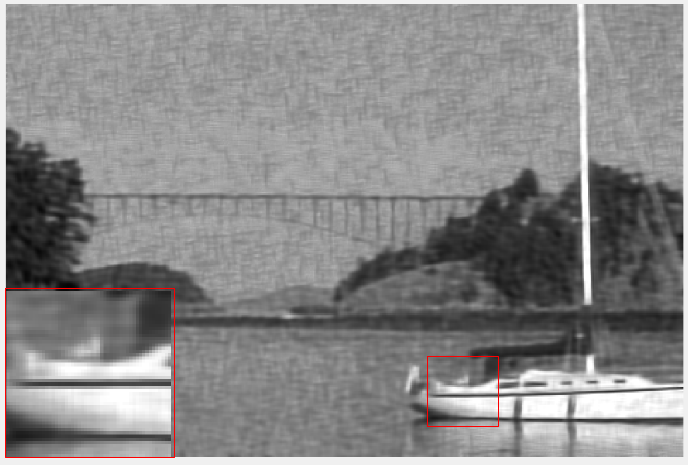}
         \caption{NL-PCA}
         \label{fig:river_pca}
     \end{subfigure}
          \begin{subfigure}[b]{0.16\textwidth}
         \centering
         \includegraphics[width=\textwidth]{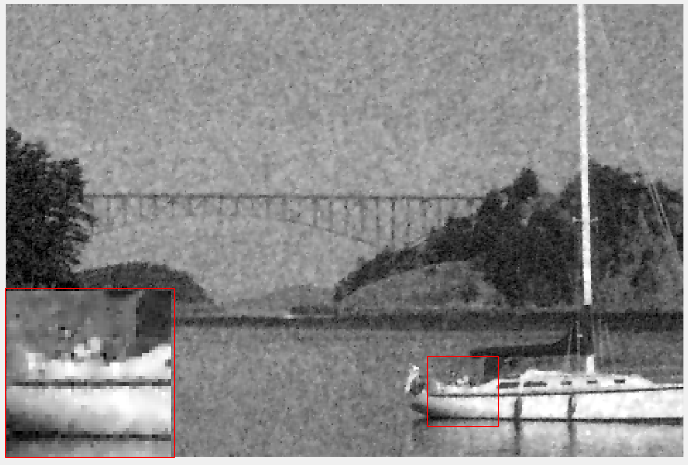}
         \caption{TV}
         \label{fig:bridge_tv}
     \end{subfigure}
          \begin{subfigure}[b]{0.16\textwidth}
         \centering
         \includegraphics[width=\textwidth]{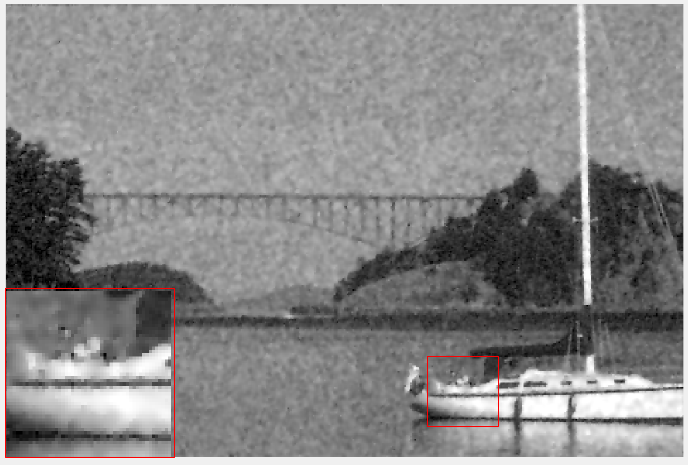}
         \caption{FOTV}
         \label{fig:bridge_fotv}
     \end{subfigure}
               \begin{subfigure}[b]{0.16\textwidth}
         \centering
         \includegraphics[width=\textwidth]{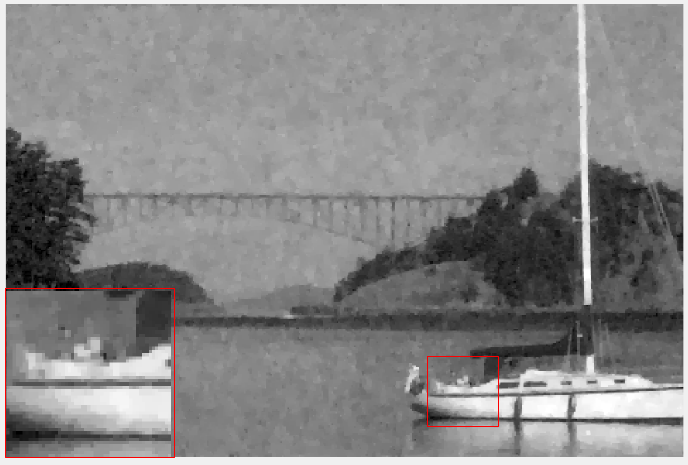}
         \caption{AITV}
         \label{fig:bridge_aitv}
     \end{subfigure}
        \caption{Poisson denoising results for Figure \ref{fig:river} with peak value 30.}
        \label{fig:close_up}
\end{figure*}
\begin{figure}
    \centering
    \includegraphics[scale=0.3]{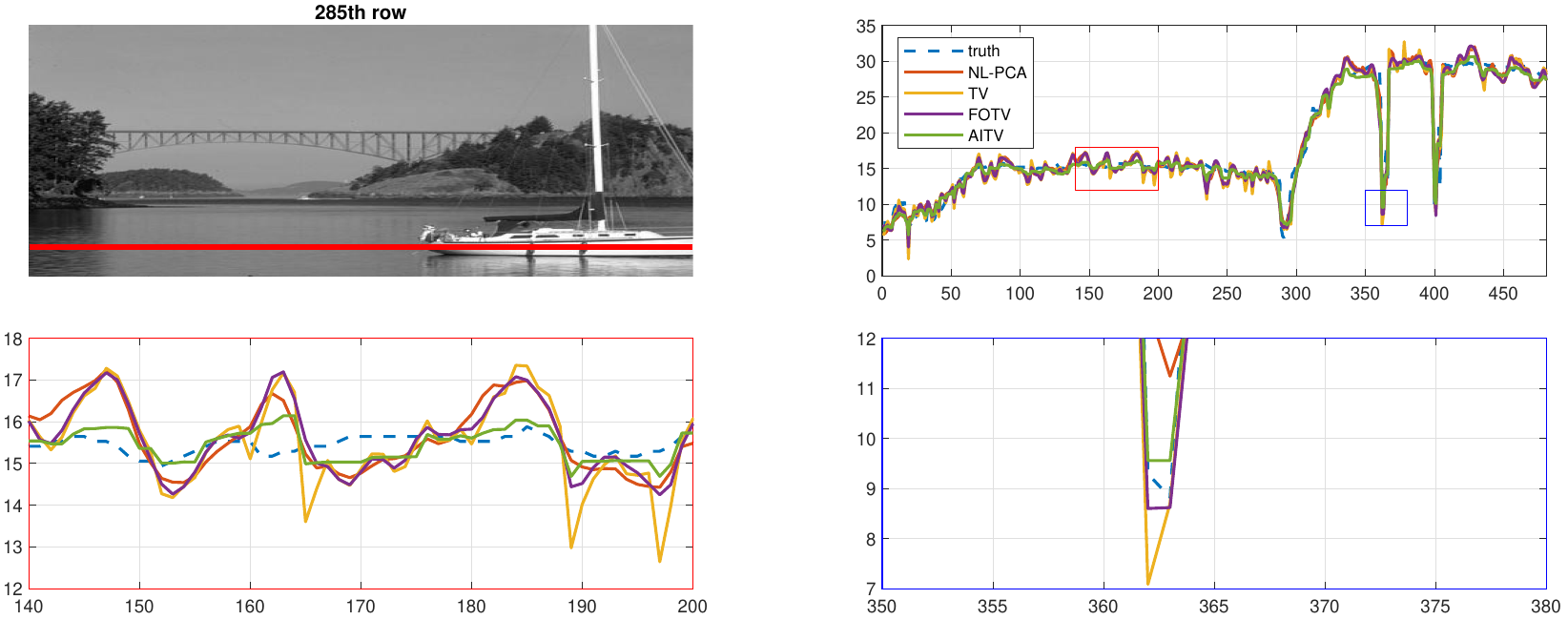}
    \caption{Reconstructed line profiles from Figure \ref{fig:close_up} over the ground truth line profile for the 285th row of Figure \ref{fig:river}. }
    \label{fig:profile}
\end{figure}

The parameters for each method are carefully tuned for the best PSNR. As the TV, FOTV, and AITV models are solved by ADMM, their parameters are nearly the same. We fix the penalty parameter $\beta = 10^{-3}$ and find the optimal fidelity parameter $\lambda$ from $\{3, 5, 8, 10, 12, 15, 20\}$. The fractional order for FOTV is optimized among $\{1.2, 1.4, 1.6, 1.8\}$.  The AITV parameter $\alpha$ is optimized among $\{0.1i\}_{i=1}^5$. We fix $\sigma = 1.75$ in the proposed Algorithm~\ref{alg:admm}. The stopping conditions are  up to 300 iterations with a relative error stopping criterion $\frac{\|u_{k}-u_{k-1}\|_2}{\|u_{k}\|_2} < 10^{-5}$. For NL-PCA, we tune the patch size and the number of clusters, which are selected from $\{3,5,7,9\}$ and $\{15, 20, 25\}$, respectively. 

Since Poisson noise depends on the pixel intensity, we control the noise level by changing the peak value of an image. In particular, before adding Poisson noise to an image, we rescale its peak value to 80, 55, and 30 with a lower peak corresponding to a noisier image. Table \ref{tab:result} records the PSNR and SSIM metrics of the denoised images by various competing methods, showing that the proposed AITV model achieves the best results in most cases. The average results are also shown in the last column, showcasing that AITV is consistently the best under the three peak values. 

We examine Figure \ref{fig:river}. The image has peak value 30 before adding Poisson noise. The denoised results  are presented in Figure \ref{fig:close_up}. By focusing on the sky, the denoised image by AITV looks less noisy compared to the other methods. From the enlarged  window with red boundary, the result appears sharper for AITV than the other methods. Figure \ref{fig:profile} compares a line profile from a denoised image with its original, showing that AITV is the least noisy.

Table \ref{tab:time} reports the average computational time over all the testing scenarios (any combination of five images and three peak levels), demonstrating that the proposed  Algorithm \ref{alg:admm} is nearly one order of magnitude (2 seconds) faster than the other methods (about 20 seconds). Overall, the proposed AITV model solved by our designed ADMM algorithm is the most efficient in computation time and most effective in PSNR and SSIM.

\section{Conclusion} \label{sec:conclusion}
In this work, we formulated a variational Poisson model with AITV regularization. To solve the model, we designed an efficient ADMM incorporating the $\ell_1 - \alpha \ell_2$ proximal operator that numerically converges within seconds. Our experiments demonstrated the efficiency of  the proposed approach over several representative Poisson denoising methods in terms of quantitative measures, visual appearance, and computational time. For  future directions, we will extend the proposed model to color images and design a globally convergent algorithm. Extending to deep learning, we will unfold the proposed ADMM algorithm in a similar fashion as \cite{zheng2021adaptive}.

\bibliographystyle{IEEEbib-abbrev}\bibliography{test}

\end{document}